\documentstyle[psfig]{l-aa}     

\def\ros{{\sl ROSAT}}
\def\etal{{et\,al. }}
\def\iue{{\sl IUE}}
\def\exo{{\sl EXOSAT}}
\def\ein{{\sl Einstein}}

\def\msun{M$_{\odot}$}
\def\mt{$\tilde{m}$}
\def\rsun{R$_{\odot}$}
\def\lsun{L$_{\odot}$}
\def\mdot{$\dot M$}
\def\grad{$^\circ$}
\def\it{\sl}
\def\degs{\ifmmode ^{\circ}\else$^{\circ}$\fi}
\def\amin{\ifmmode ^{\prime}\else$^{\prime}$\fi}
\def\asec{\ifmmode ^{\prime\prime}\else$^{\prime\prime}$\fi}
\def\fm{\hbox{$.\!\!^{\rm m}$}}            
\def\fss{\hbox{$.\!\!^{\rm s}$}}        
\def\farcs{\hbox{$.\!\!^{\prime\prime}$}}  
\newbox\grsign \setbox\grsign=\hbox{$>$}
\newdimen\grdimen \grdimen=\ht\grsign
\newbox\laxbox \newbox\gaxbox
\setbox\gaxbox=\hbox{\raise.5ex\hbox{$>$}\llap
     {\lower.5ex\hbox{$\sim$}}}\ht1=\grdimen\dp1=0pt
\setbox\laxbox=\hbox{\raise.5ex\hbox{$<$}\llap
     {\lower.5ex\hbox{$\sim$}}}\ht2=\grdimen\dp2=0pt
\def\gax{$\mathrel{\copy\gaxbox}$}
\def\lax{$\mathrel{\copy\laxbox}$}
\def\h{$^{\rm h}$}\def\m{$^{\rm m}$}

\def\ag{AG\,Dra}
\font\pr=cmr7

\def\II{{\pr II}}
\def\III{{\pr III}}
\def\IV{{\pr IV}}
\def\V{{\pr V}}

\begin{document}
 
   \thesaurus{06         
               02.01.2;  
               08.09.2 (AG Dra);  
               08.14.2;  
               08.22.3;  
               13.21.5;  
               13.25.5   
               }

   \title{The UV/X-ray emission of the symbiotic star AG Draconis \\
   during quiescence and the 1994/1995 outbursts\thanks{$\!\!\!$Based on 
    observations made with the
  {\it International Ultra\-violet Explor\-er} collected at the
Villa\-franca Satellite Tracking Station of the European Space Agency,
Spain.}}
 
   \author{J. Greiner\inst{1}, K. Bickert\inst{1}, R. Luthardt\inst{2}, 
      R. Viotti\inst{3}, A. Altamore\inst{4}, 
      R. Gonz\'alez-Riestra\inst{5}\thanks{$\!\!\!$Also Astronomy Division, 
      Space Science Department, ESTEC.},
      R.E. Stencel\inst{6}}

   \offprints{J.\,Greiner,\,jcg@mpe-garching.mpg.de}
 
  \institute{$^1$ Max-Planck-Institut f\"ur extraterrestrische Physik,
             85740 Garching, Germany \\
         $^2$ Sternwarte Sonneberg, 96515 Sonneberg, Germany \\
        $^3$ Istituto di Astrofisica Spaziale, CNR, Via Enrico Fermi 21, 
              00044 Frascati, Italy \\
        $^4$ Dipartimento di Fisica E. Amaldi, Universit\`a Roma III, 
             00146 Roma, Italy \\
        $^5$ IUE Observatory, ESA, Villafranca del Castillo, P.O. Box 50727,
             28080 Madrid, Spain \\
        $^6$ Department of Physics and Astronomy, University of Denver, 
             Denver, CO 80208, USA }

   \date{Received  May 3, 1996; accepted November 22, 1996}
 
   \maketitle
   \markboth{Greiner et al.}{UV/X-ray emission of AG Dra}
 
   \begin{abstract}
We present the results of an extensive campaign of coordinated X-ray (\ros) and
UV (\iue) observations of the symbiotic star AG Dra during a long period 
of quiescence followed recently by a remarkable
phase of activity characterized by two optical outbursts.
The major optical outburst in June 1994 and
the secondary outburst in July 1995 were covered by a number of target
of opportunity observations (TOO) with both satellites. Optical photometry
is used to establish the state of evolution along the outburst.

Our outburst observations are supplemented by a substantial number of X-ray 
observations of AG Dra during its quiescent phase between 1990--1993.
Near-simultaneous \iue\, observations at the end of 1992 are used to derive 
the spectral energy distribution from the optical to the X-ray range.
The X-ray flux remained constant over this three year quiescent phase.
The hot component (i.e. X-ray emitting compact object) turns out to be very 
luminous: a blackbody fit to the
X-ray data in quiescence with an absorbing column equal to the total 
galactic N$_{\rm H}$ in this direction gives
(9.5$\pm$1.5)$\times$10$^{36}$ (D/2.5 kpc)$^2$ erg/s. This suggests that the 
compact object is burning hydrogen-rich matter on its surface even in the 
quiescent (as defined optically) state at a rate of 
(3.2$\pm$0.5)$\times$10$^{-8}$ (D/2.5 kpc)$^2$ \msun/yr. Assuming a steady 
state, i.e. burning at precisely the accretion supply rate, this high rate 
suggests a Roche lobe filling cool companion though Bondi-Hoyle accretion from 
the companion wind cannot be excluded.

With \ros\, observations we have discovered a remarkable decrease of
the X-ray flux during both optical maxima, followed by a
gradual recovering to the pre-outburst flux. In the UV
these events were characterized by a large increase
of the emission line and continuum fluxes,
comparable to the behaviour of AG Dra during the 1980-81 active phase.
The anticorrelation of X-ray/UV flux and optical brightness evolution
is very likely due to a temperature decrease of the hot
component. Such a temperature decrease could be the result of an increased 
mass transfer to the burning compact object, causing it to slowly expand 
to about twice its original size during each optical outburst.


   \end{abstract}
 
\section{Introduction}

Symbiotic stars are characterized by the simultaneous occurrence in an
apparently single object of two temperature regimes differing by a factor
of 30 or more. The spectrum of a symbiotic star consists of a
late type (M-) absorption spectrum, highly excited emission lines
and a blue continuum. Generally, symbiotic stars are interpreted as interacting
binary systems consisting of a cool luminous visual primary 
and a hot compact object (white dwarf, subdwarf) as secondary
component. Because of mass loss of the giant there is often a common
nebulous envelope. Mass transfer from the cool to the hot component
is expected. An extensive review on the properties of symbiotic stars 
can be found in Kenyon (1986).

Many symbiotic stars show outburst events at optical and UV wavelengths.
Though several models have been proposed to explain these outbursts,
it is the combination of quiescent properties and these outburst properties
which make it difficult to derive a consistent picture for some symbiotics.

The symbiotic star AG Draconis (BD +67\grad922) plays an outstanding role
inside this group of stars because of its high galactic latitude,
its large radial velocity of v$_{\rm r}$=--148 km/s
and its relatively early spectral type (K).
AG Dra is probably a metal poor symbiotic binary in the galactic halo.

Here, we use all available \ros\, data to document the X-ray light curve 
of AG Dra over the past 5 years. In addition, we report on the results of the 
coordinated \ros/\iue\, campaign during the 1994/1995 outbursts.
Preliminary results on the \iue\, and optical observations were given by 
Viotti \etal (1994a, 1994b). After a description of the relevant previous 
knowledge of the AG Dra properties in 
the optical, UV and X-ray range in the remaining part of this paragraph,
we present our observational results in paragraph 2--4, discuss the various
implications on the AG Dra system parameters in paragraph 5 and end with a 
summary in paragraph 6.

\subsection{AG Dra: The optical picture}

Like in most symbiotic stars, the historical light curve of AG Dra
is characterized by a sequence of active and quiescent phases
(e.g. Robinson 1969, Viotti 1993). The activity is represented by 1--2 mag
light maxima (currently called {\it outbursts} or {\it eruptions}) frequently
followed by one or more secondary maxima.
It has been noted (Robinson 1969, Iijima \etal 1987) that the major
outbursts occur in $\approx$15 yr intervals. This recurrence period
has continued with the recent major outbursts of 1981 and
1994. Iijima \etal  (1987) suggest that the major outbursts seem to occur
at about the same orbital phase, shortly after the photometric maximum, i.e.
shortly after the spectroscopic
conjunction of the companion (hot component in front of the cool component).
Occasionally, AG Dra also undergoes smaller amplitude outbursts,
such as those in February 1985 and January 1986.

Between the active phases AG Dra is spending long periods
(few years to decades) at minimum light (V$\approx$9.8-10.0, Mattei 1995),
with small (0.1 mag) semiregular photometric variations in B and V with
pseudo-periods of 300-400 days (Luthardt 1990).
However, in the U band regular variations with amplitudes of 1 mag
and a period of 554 days have been discovered by Meinunger (1979), and
confirmed by later observations (e.g. Kaler 1987, Hric \etal 1994,
Skopal 1994).
This periodicity is associated with the orbital motion of the system,
as confirmed by the radial velocity observations of Kenyon \& Garcia
(1986), Mikolajewska \etal (1995) and Smith \etal (1996). 
Because of the color dependence these U band variations certainly 
reflect the modulation of the Balmer continuum emission.
The photometric variability is continuous, suggesting periodic
eclipses of an extended hot region by the red giant.

In June 1994 Graslo \etal (1994) announced that AG Dra was starting
a new active phase which was marked by a rapid brightening from V=9.9
to V=8.4 on June 14th, and to 8.1 on July 6-10, 1994.
After July 1994 the brightness gradually declined reaching the quiescent level
(V$\approx$9.8) in November 1994.
Like the 1981--82 and 1985--86 episodes, AG Dra underwent a
secondary outburst in July 1995, reaching the light maximum of V=8.9 by
the end of the month.

The optical spectrum of AG Dra was largely investigated especially in
recent years, during both the active phases and quiescence.
The spectrum is typical of a symbiotic star, with a probably stable
cool component which dominates the yellow-red region, and a largely
variable ``nebular" component with a strong blue-ultraviolet continuum and
a rich emission line spectrum (e.g. Boyarchuk 1966).
According to most authors the cool component is a K3 giant,
which together with its large radial velocity (--148 km s$^{-1}$) and
high galactic latitude (b$^{\rm II}$=+41$^{\circ}$), would place AG Dra in
the halo population at a distance of about 1.2 kpc, 0.8 kpc above the galactic 
plane. While Huang \etal (1994) suggested that the cool component of AG Dra 
may be of spectral type K0Ib, which would place it at a distance of about 
10 kpc,
Mikolajewska \etal (1995) from a comparison with the near-infrared colors 
of M3 and M13 giants concluded that the cool component is a bright giant with
M$_{\rm bol}$ \gax --3.5, placing AG Dra at about 2.5 kpc.
Most recently, Smith \etal (1996) performed a detailed abundance analysis
and found a low metallicity ([Fe/H]=--1.3), a temperature of
T$_{\rm eff}$=4300$\pm$100 K and log g = 1.6$\pm$0.3 consistent with the 
classification of AG Dra's cool component as an early K giant.
In the following we shall assume a distance of 2.5 kpc for the AG Dra system.

\subsection{AG Dra: The UV picture}

The hot dwarf companion is a
source of intense ultraviolet radiation which produces a rich
high-temperature emission line spectrum and a strong UV continuum.
AG Dra is in fact a very bright UV target which has been intensively
studied with \iue\, (e.g. Viotti \etal 1983, Lutz \etal 1987,
Kafatos \etal 1993, M\"urset \etal 1991, Mikolajewska \etal 1995).

Ultraviolet high-resolution spectra with the \iue\, satellite revealed 
high-ionization permitted emission lines such as the resonance doublets of 
N{\V}, C{\IV} and Si\IV. The strongest emission line is 
He{\II} $\lambda$1640 which is 
composed of narrow and broad (FWHM = 6 \AA, or equivalently $\approx$1000 km/s)
components. The N{\V} line has a P Cygni absorption component displaced
120 km/s from the emission peak (Viotti \etal 1983).

The origin of this feature is unclear. It can either arise in the red-giant
wind ionized by the hot dwarf radiation, or in some low-velocity regions of
the hot component's wind.

The UV continuum and line flux is largely variable with the star's activity.
Viotti \etal (1984) studied the IUE spectra of AG Dra during the
major 1980--1983 active phase, and found that the outburst was most
energetic in the ultraviolet with an overall rise of about a factor 10
in the continuum, much larger than in the visual,
and of a factor 2--5 in the emission line flux.
A large UV variation was also a characteristics of the minor
1985 and 1986 outbursts (e.g. Mikolajewska \etal 1995).

\subsection{AG Dra: The X-ray picture}

First X-ray observations of AG Dra during the quiescent phase were done 
with the HEAO-2 satellite (\ein\, Observatory) before the 1981-1985 series of
eruptions (0.27 IPC counts/sec). The spectrum was found to be very soft 
(Anderson \etal 1981). The data are consistent
with a blackbody source of kT=0.016 keV (Kenyon 1988) in addition to the
bremsstrahlung source (kT=0.1 keV) suggested by Anderson \etal (1981).
The X-ray temperature of $\approx$~200~000 K is in fair agreement with the
source temperature of  $\approx$\,100\,000--150\,000~K inferred from \iue\,
observations, although the blackbody radius deduced from \iue\, data is
nearly a factor of 10 larger than the X-ray value (Kenyon 1988).
Unfortunately, the major 1980 outburst was ``lost" by the HEAO-2
satellite because of a failure of the high voltage power.

A comparison of the X-ray flux with the observed He{\II} $\lambda$4686 flux 
indicates that the luminosity in absorbed He$^+$ photons is larger 
by a factor of two (Kenyon 1988). It has been concluded that the
X-rays are degraded by the surrounding nebula thus causing an underestimate 
of the actual X-ray luminosity.

\exo\, was pointed on AG Dra four times during the 1985--86 minor active phase,
which was characterized by two light maxima in February 1985
and January 1986. These observations revealed a large X-ray fading with 
respect to quiescence (Piro 1986), the source being at least 5--6 times 
weaker in the EXOSAT thin Lexan filter in March 1985, and not detected in 
February 1986 (Viotti \etal 1995).
Simultaneous IUE observations have on the contrary shown an increase
of the continuum and emission line flux, especially of the high
temperature N{\V} 1240 A and He{\II} 1640 A lines at the time of the
light maxima. According to Friedjung (1988) this behaviour might be due to 
a temperature drop of a non-black body component, or to a continuous absorption
of the X-rays shortwards of the N$^{+3}$ ionization limit.
No eclipse of the X-ray source was found at phase 0.5 (beginning of
November 1985) of the orbital motion of the AG Dra system,
implying a limit in the orbit inclination.
The weakness of the countrate in the Boron (filter \#6) EXOSAT filter during
quiescence implies a low (2--3$\times$10$^5$ K) temperature of the source
(Piro \etal 1985).

   \begin{table}
     \vspace{-0.25cm}
      \caption{Photometric observations of AG Dra at Sonneberg Observatory
           (see paragraph 2 for details).}
           \vspace{-0.4cm}
            \begin{center}
             \begin{tabular}{cccrcr}
            \hline
            \noalign{\smallskip}
   HJD &     V  & HJD  &   B~~   & HJD & U~~\\
       &$\!$(mag)$\!$ &      &$\!$(mag)$\!$ &     & $\!$(mag)$\!$\\
            \noalign{\smallskip}
            \hline
            \noalign{\smallskip}
8127.3876 & 9.76 & 8127.3913 & 11.15 & 8127.3991 & 11.77 \\
8179.3537 & 9.85 & 8179.3592 & 11.25 &  &  \\
8179.3537 & 9.85 & 8179.3613 & 11.26 & 8179.3696 & 11.66 \\
8274.6330 & 9.83 & 8274.6384 & 11.20 & 8274.6432 & 11.37 \\
8329.6163 & 9.76 & 8329.6212 & 11.10 & 8329.6265 & 11.32 \\
8332.5781 & 9.76 & 8332.5840 & 11.11 & 8332.5908 & 11.34 \\
8353.5254 & 9.69 & 8353.5317 & 11.06 & 8353.5380 & 11.36 \\
8359.5031 & 9.70 & 8359.5076 & 11.07 & 8359.5135 & 11.31 \\
8362.5344 & 9.72 & 8362.5425 & 11.07 & 8362.5498 & 11.33 \\
8409.4444 & 9.73 & 8409.4491 & 11.08 & 8409.4544 & 11.28 \\
8440.4711 & 9.71 & 8440.4755 & 11.09 & 8440.4809 & 11.43 \\
8494.3874 & 9.77 & 8494.3916 & 11.15 & 8494.3978 & 11.67 \\
8500.4155 & 9.74 & 8500.4208 & 11.11 & 8500.4275 & 11.64 \\
8681.4692 & 9.71 & 8681.4732 & 11.08 & 8681.4785 & 11.54 \\
8691.6060 & 9.73 & 8691.6099 & 11.08 & 8691.6153 & 11.59 \\
8747.4630 & 9.72 & 8747.4699 & 11.02 & 8747.4767 & 11.33 \\
8801.4260 & 9.75 & 8801.4301 & 11.08 & 8801.4362 & 11.35 \\
8803.4410 & 9.74 & 8803.4480 & 11.08 & 8803.4540 & 11.33 \\
8830.3997 & 9.85 & 8830.4040 & 11.19 & 8830.4107 & 11.41 \\
8843.3743 & 9.83 & 8843.3781 & 11.18 & 8843.3831 & 11.31 \\
9213.3667 & 9.41 & 9213.3746 & 10.40 & 9213.3818 &  9.89 \\
9214.4254 & 9.48 & 9214.4333 & 10.53 & 9214.0000 & 10.42 \\
9215.3892 & 9.54 & 9215.3981 & 10.62 & 9215.4061 & 10.28 \\
9217.3517 & 9.49 & 9217.3587 & 10.57 & 9217.3648 & 10.15 \\
9249.3329 & 9.63 & 9249.3385 & 10.83 & 9249.3448 & 10.56 \\
9250.3517 & 9.64 & 9250.3581 & 10.82 & 9250.3639 & 10.54 \\
9534.4792 & 8.89 & 9534.4863 &  9.56 & 9534.4931 &  8.60 \\
9535.4286 & 8.77 & 9535.4350 &  9.34 & 9535.4419 &  8.43 \\
9537.4377 & 8.67 & 9537.4436 &  9.22 & 9537.4494 &  8.32 \\
9541.4588 & 8.46 & 9541.4740 &  8.96 & 9541.4790 &  8.04 \\
9545.4204 & 8.55 & 9545.4266 &  9.02 & 9545.4327 &  8.15 \\
9568.4060 & 8.61 & 9568.4117 &  9.09 & 9568.4182 &  8.30 \\
9569.4107 & 8.62 & 9569.4172 &  9.11 & 9569.4229 &  8.28 \\
9599.3582 & 8.85 & 9599.3688 &  9.50 & 9599.3760 &  8.72 \\
9623.4056 & 9.06 & 9623.4124 &  9.78 & 9623.4187 &  9.07 \\
9625.4310 & 9.07 & 9625.4376 &  9.80 & 9625.4438 &  9.09 \\
9638.3671 & 9.17 & 9638.3732 &  9.92 & 9638.3791 &  9.19 \\
9639.3749 & 9.15 & 9639.3824 &  9.96 & 9639.3885 &  9.25 \\
9644.4397 & 9.19 & 9644.4462 &  9.97 & 9644.4544 &  9.20 \\
            \noalign{\smallskip}
           \hline
           \end{tabular}
           \end{center}
           \label{phot}
           \vspace{-.35cm}
   \end{table}

   \begin{table}
        \vspace{-0.25cm}
      \caption{Photographic observations of AG Dra at Sonneberg Observatory
           (see paragraph 2 for details).}
           \vspace{-0.4cm}
            \begin{center}
             \begin{tabular}{cccr}
            \hline
            \noalign{\smallskip}
   HJD &     m$_{\rm pg}$  & HJD  &   m$_{\rm pg}$~   \\
       & (mag) &      & (mag)$\!$ \\
            \noalign{\smallskip}
            \hline
            \noalign{\smallskip}
9002.317 &  10.6 & 9476.471 &  10.6 \\
9005.310 &  10.6 & 9476.491 &  10.5 \\
9027.258 &  10.6 & 9476.491 &  10.4 \\
9028.259 &  10.5 & 9480.480 &  10.6 \\
9029.263 &  10.7 & 9480.480 &  10.5 \\
9030.298 &  10.7 & 9480.480 &  10.5 \\
9031.303 &  10.5 & 9481.488 &  10.5 \\
9032.300 &  10.8 & 9481.488 &  10.4 \\
9041.276 &  10.8 & 9481.488 &  10.6 \\
9066.531 &  11.0 & 9482.487 &  10.5 \\
9094.472 &  11.0 & 9482.487 &  10.5 \\
9094.472 &  10.5 & 9482.487 &  10.4 \\
9098.500 &  11.0 & 9484.494 &  10.4 \\
9098.500 &  10.7 & 9484.494 &  10.4 \\
9099.486 &  11.0 & 9484.494 &  10.5 \\
9099.486 &  10.9 & 9486.456 &  10.5 \\
9101.507 &  11.0 & 9486.456 &  10.3 \\
9101.507 &  10.9 & 9486.456 &  10.4 \\
9154.458 &  11.0 & 9488.493 &  10.5 \\
9154.458 &  10.9 & 9488.493 &  10.5 \\
9249.496 &  10.8 & 9488.493 &  10.5 \\
9250.487 &  10.6 & 9504.419 &   9.4 \\
9278.519 &  10.5 & 9511.453 &   9.2 \\
9279.468 &  10.5 & 9518.458 &   9.1 \\
9310.486 &  10.7 & 9535.417 &   9.0 \\
9416.304 &  10.5 & 9536.422 &   9.0 \\
9422.549 &  10.7 & 9537.423 &   8.9 \\
9422.549 &  10.7 & 9541.415 &   8.9 \\
9422.549 &  10.6 & 9574.422 &   8.7 \\
9457.489 &  10.5 & 9836.477 &  10.0 \\
9457.499 &  10.5 & 9839.479 &  10.0 \\
9457.499 &  10.5 & 9840.483 &  10.2 \\
9458.517 &  10.5 & 9841.511 &  10.3 \\
9458.517 &  10.4 & 9842.522 &  10.1 \\
9458.517 &  10.6 & 9862.443 &  10.1 \\
9462.550 &  10.5 & 9888.422 &  10.3 \\
9462.550 &  10.4 & 9889.422 &  10.0 \\
9462.550 &  10.6 & 9894.422 &   9.9 \\
9475.478 &  10.5 & 9895.422 &   9.8 \\
9475.478 &  10.4 & 9896.417 &   9.9 \\
9475.478 &  10.6 &          &        \\
            \noalign{\smallskip}
           \hline
           \end{tabular}
           \end{center}
           \label{shue}
           \vspace{-.45cm}
   \end{table}

\section{Optical observations}

AG Dra is included in a long-term programme of optical photometry 
at the 60 cm Cassegrain telescope of Sonneberg Observatory. 
A computer controlled, photoelectric photometer with a diaphragm of 20\asec\, 
diameter  is used to obtain consecutive UBV images of AG Dra, the comparison
star BD+67 952 and a third star (SAO 16935) used for check purposes.
Typical integration times were 20 sec and the brightnesses were reduced to the
international system according to Johnson. The mean errors in the brightness
determination are $\pm$0.01 mag in V and B and $\pm$0.03 mag in U.
The photometric observations since June 1990 (for which near-simultaneous 
X-ray measurements with \ros\, are available) are tabulated in Tab. \ref{phot} 
and the U and B magnitudes are also plotted in Fig. \ref{light}. Earlier
Sonneberg photometric data of AG Dra are included in Hric \etal (1993).

In addition to the photoelectric data we have also used the photographic
sky patrol of Sonneberg Observatory to fill up gaps in the 
photoelectric coverage. The sky patrol is performed simultaneously in 
two colors (red and blue). The brightness measurements of AG Dra from the 
blue plates covering the time interval of the two optical outbursts
are given in Tab. \ref{shue} and are also plotted in the top panel of
Fig. \ref{light}.

\section{IUE observations}

{\it IUE} observations were mostly obtained as a Target-of-Opport\-unity
(TOO) programme in coordination with the ROSAT observations.
The last observations were made within the approved programme SI047
for the 19th IUE observing episode. 
Tab. \ref{IUEobs} summarizes the {\it IUE} low resolution
observations made in the period June 1994 -- September 1995.

   \begin{table}
         \vspace{-0.25cm}
      \caption{Journal of AG Dra low resolution \iue\, observations}
            \begin{center}
             \begin{tabular}{lcccc}
            \hline
            \noalign{\smallskip}
Image$^a$ & Disp & Aper$^b$ & Date       &  T$_{\rm exp}^c$   \\
      &      &      & dd/mm/yy   &  min:sec         \\
            \noalign{\smallskip}
            \hline
            \noalign{\smallskip}
SWP51315 & L & L,S & 04/07/94 &  15:00,5:00  \\
LWP28542 & L & L,S & 04/07/94 &   7:00,3:00  \\
SWP51331 & L & L,S & 07/07/94 &   2:00,2:00  \\
LWP28556 & L & L,S & 07/07/94 &   1:30,1:00  \\
SWP51415 & L & L,S & 12/07/94 &   1:00,1:00  \\
LWP28628 & L & L,S & 12/07/94 &   1:00,1:00  \\
SWP51416 & L & L,S & 12/07/94 &   1:35,1;35  \\
SWP51632 & L & L,S & 28/07/94 &   2:00,1:00  \\
LWP28752 & L & L,S & 28/07/94 &   2:00,1:00  \\
SWP52143 & L & L   & 19/09/94 &   1:30       \\
LWP29192 & L & L   & 19/09/94 &   1:00       \\
SWP52145 & L & L   & 19/09/94 &   3:00       \\
SWP52994 & L & L,S & 06/12/94 &   3:00,2:00  \\
LWP29654 & L & L,S & 06/12/94 &   2:00,2:00  \\
SWP54632 & L & L,S & 08/05/95 &   3:00,1:30  \\
LWP30642 & L & L,S & 08/05/95 &   2:00,1:00  \\
SWP55371 & L & L,S & 28/07/95 &   1:00,1:00  \\
LWP31175 & L & L,S & 28/07/95 &   2:00,1:00  \\
SWP55929 & L & L,S & 14/09/95 &   2:00,1:00  \\
LWP31476 & L & L,S & 14/09/95 &   3:00,1:30  \\
            \noalign{\smallskip}
           \hline
           \end{tabular}
           \end{center}
           \label{IUEobs}
           \vspace{-.2cm}

   \noindent{\small 
         $^a$ $SWP$: 1200-1950 \AA. $LWP$: 1950-3200 \AA. \\
         $^b$ $L, S$: both large and small \iue\, apertures were
              used. $L$: only the large aperture was used. \\
         $^c$ Two exposure times correspond to Large and Small aperture,
               respectively. }
   \vspace{-0.2cm}
   \end{table}

   \begin{table}
      \vspace{-0.25cm}
      \caption{Ultraviolet fluxes$^a$ of AG Dra in outburst. Typical
           quiescent fluxes are given in the first line (Viotti \etal 
            1984, Meier \etal 1994).}
           \vspace{-0.4cm}
            \begin{center}
            \begin{tabular}{cclrccr}
            \hline
            \noalign{\smallskip}
      Date$^b$ & N\V & ~1400$^c$ & C\IV  & He\II & F$_{1340}$ & 
                F$_{2860}$ \\
            \noalign{\smallskip}
            \hline
            \noalign{\smallskip}
    quiescence & ~~10 & ~~~\,\,8     &~~~~9 &~~~26 &~~~29 &~~~12 \\
      9537 & ~~52 &~~~32     &~~114 &~~237 &~~607 &~~368 \\
      9538 & ~~41 &~~~26     &~~~99 &~~277 &~~535 &~~367 \\
      9545 & ~~35 &~~~25     &~~~80 &~~221 &~~656 &~~407 \\
      9560 & ~~19 &~~~17     &~~~49 &~~197 &~~526 &~~303 \\
      9612 & ~~39 &~~~23     &~~~50 &~~150 &~~299 &~~191 \\
      9692 & ~~23 &~~~11     &~~~32 &~~121 &~~166 &~~~83 \\
      9844 & ~~29 &~~~11     &~~~25 &~~195 &~~144 &~~~64 \\
      9926 & ~~36 &~~~61$^d$ &~~~59 &~~183 &~~137 &~~158 \\
      9975 & ~~36 &~~~31     &~~~68 &~~172 &~~155 &~~~55 \\
            \noalign{\smallskip}
           \hline
           \end{tabular}
           \end{center}
        \vspace{-0.2cm}

\noindent{\small 
$^a$ Emission line (in 10$^{-12}$ erg cm$^{-2}$ s$^{-1}$) and continuum
     fluxes (in 10$^{-14}$ erg cm$^{-2}$ s$^{-1}$ \AA$^{-1}$),
     dereddened for E$_{\rm B-V}$=0.06. \\
$^b$ JD -- 2440000. \\
$^c$ Blend of O \IV] and Si \IV. O \IV] $\lambda$1401.16 is
     the main contri\-butor to the blend. \\
$^d$ Broad feature.}
           \label{iueflux}
           \vspace{-.2cm}
   \end{table}

   \begin{figure*}[htbp]
      \centering{
      \hspace*{.1cm}
      \vbox{\psfig{figure=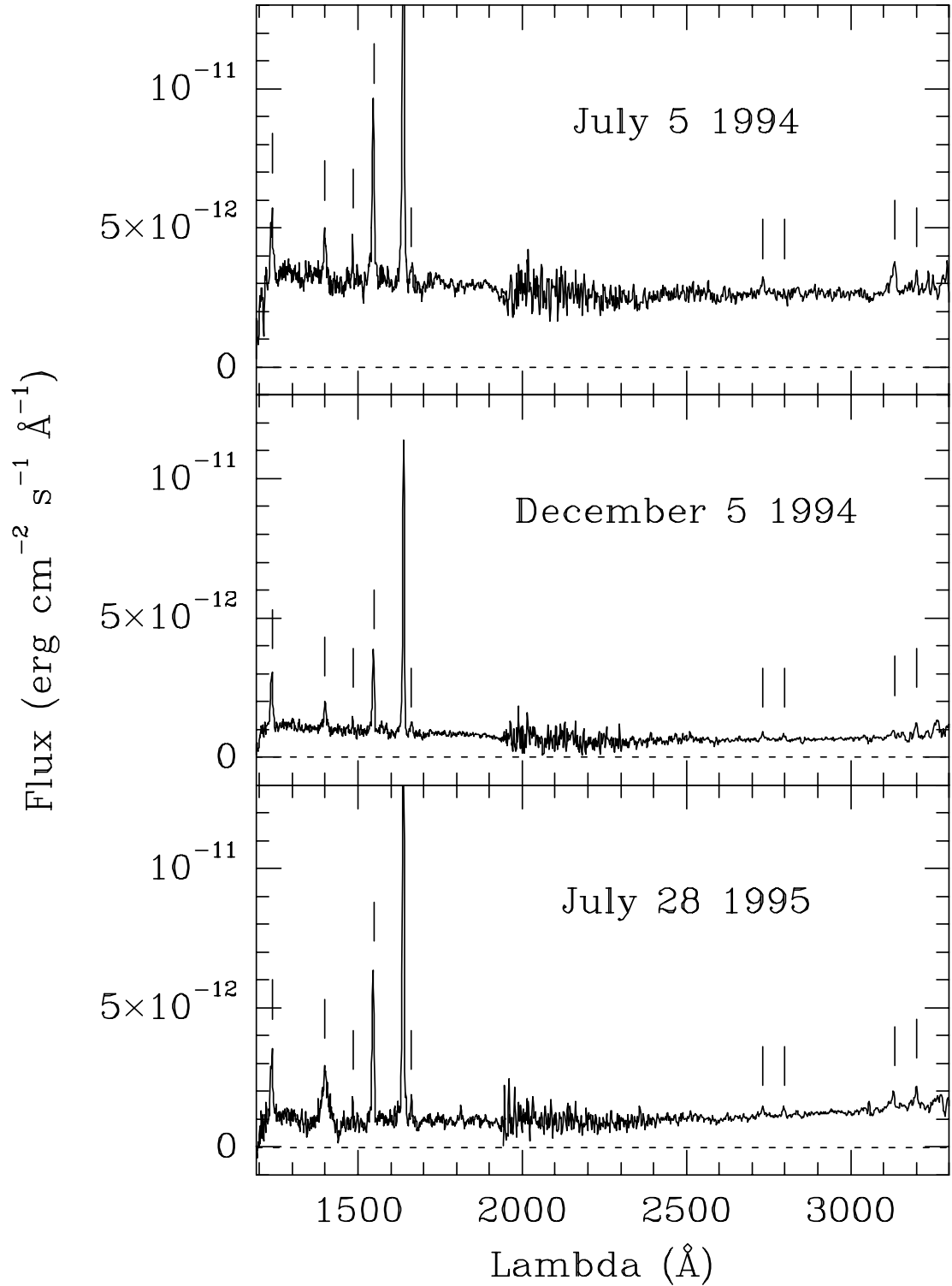,width=15cm,%
          bbllx=4.6cm,bblly=9.cm,bburx=13.2cm,bbury=20.6cm,clip=}}\par
      }
      \caption[iue2]{IUE spectra of AG Dra at three different dates during 
          the 1994-1995 activity phase: during the 1994 outburst (top),
          during the ``quiescent" state between the outbursts (middle)
          and during the second outburst (bottom). 
          The strongest line in the three spectra is He{\II} 1640 \AA.
          The lines marked in the spectra are: N{\V} 1240 \AA,
          the blend Si{\IV} + O\IV] 1400 \AA, 
          N{\IV}] 1486 \AA, C{\IV} 1550 \AA, O\III] 1663 \AA, 
          He{\II} 2733 \AA, Mg{\II} 2800 \AA, O{\III} 3133 \AA\ and 
          He{\II} 3202 \AA.}
         \label{iue2}
    \end{figure*}

In general, priority was given to the low resolution spectra for line and
continuum fluxes. High resolution spectra were also taken at all dates,
but they will not be discussed in this paper. For most of the low resolution
images, both IUE apertures
were used in order to have  in the Large Aperture the continuum and the
emission lines well exposed and in the Small Aperture (not photometric) the
strongest emission lines (essentially He{\II} 1640 \AA) not saturated.
The fluxes from the Small Aperture spectra were corrected for the smaller
aperture transmission by comparison with the non-saturated parts of the
Large Aperture data. Some representative IUE spectra are shown in Fig. 
\ref{iue2}.

   \begin{figure}[htbp]
      \centering{
       \vspace{0.1cm}
      \vbox{\psfig{figure=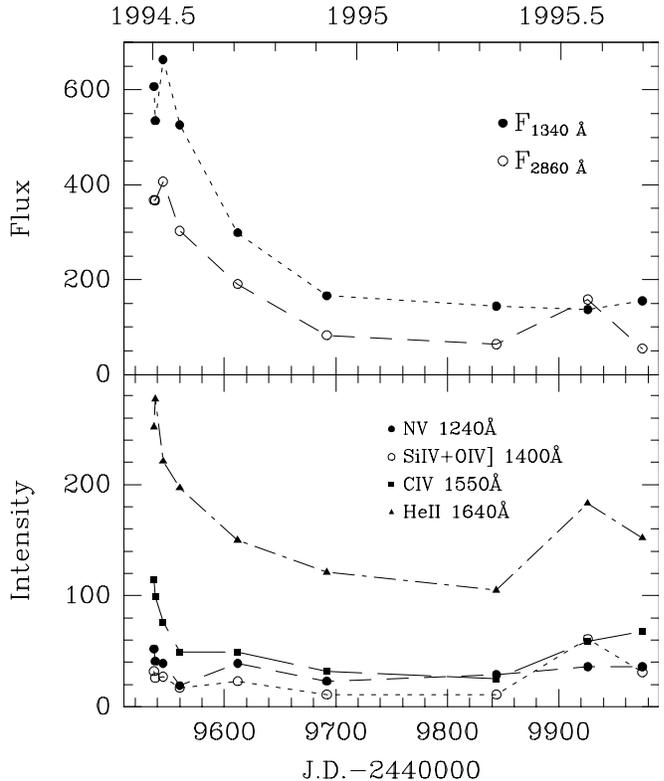,width=8.8cm,%
          bbllx=4.0cm,bblly=12.2cm,bburx=12.4cm,bbury=22.2cm,clip=}}\par}
      \vspace{-0.65cm}
      \caption[iue1]{Evolution of the UV continuum and emission lines of AG Dra 
        during the 1994-1995 outburst. Units of continuum fluxes are 
        10$^{-14}$ erg cm$^{-2}$
        s$^{-1}$ \AA$^{-1}$, and of line intensities  10$^{-12}$ erg cm$^{-2}$
        s$^{-1}$. All the measurements are corrected for an interstellar
        reddening of E(B-V)=0.06.}
         \label{iue1}
         \vspace{-0.15cm}
    \end{figure}

The flux of the strongest emission lines and of the continuum flux at 1340 and 
2860 \AA\ is given in Tab. \ref{iueflux}. All the fluxes are corrected for an
interstellar absorption of E(B-V)=0.06 (Viotti \etal 1983). The feature at
1400 \AA\ is a blend, unresolved at low resolution of the Si\IV\, doublet at
1393.73 and 1402.73 \AA, the O\IV] multiplet (1399.77, 1401.16, 1404.81 and
1407.39 \AA), and S\IV] 1402.73 \AA. O\IV] 1401.16 \AA\, is the dominant 
contributor to the blend. Because of the variable intensity of the lines, at 
low resolution the 1400 \AA\ feature shows extended wings with variable
extension and strength (Fig. \ref{iue2}). The two continuum regions were 
chosen, as in previous works, for being less affected by the emission lines, 
and for being the 1340 \AA\ representative of the continuum of the hot source
(nearly the Rayleigh-Jeans tail of a 10$^{5}$ K blackbody), and the 2860 \AA\
region representative of the H{\II} Balmer continuum emission (see for
instance Fern\'andez-Castro \etal 1995).

The behaviour of the continuum and line intensities is shown in Fig. \ref{iue1}.
In July 1994 the UV continuum flux increase was larger than that of the
He{\II} 1640 \AA\ line, suggesting  a decrease of the Zanstra temperature at
the time of the outburst, and a recovering of the high temperature in
September 1994. In addition, the N{\V}/continuum (1340 \AA) and the
He{\II}/continuum (1340 \AA) ratios reached a maximum at the time of the second
outburst (July 1995). The He{\II} Zanstra temperatures are given in 
Fig. \ref{light}.

\section{ROSAT observations}

\subsection{Observational details}

\subsubsection{All-Sky-Survey}

AG Dra was scanned during the All-Sky-Survey over a time span of 10 days. 
The total observation time resulting from 95 individual scans adds up to 
2.0 ksec. All the \ros\, data analysis described in the following has been 
performed using the dedicated EXSAS package (Zimmermann \etal 1994).

Due to the scanning mode the source has been observed at all possible off-axis
angles with its different widths of the point spread function. For the 
temporal and spectral analysis we have used an 5\amin\, extraction radius 
to ensure that no source photons are missed. No other source down to the 
1$\sigma$ level is within this area. Each photon
event has been corrected for its corresponding effective area. The background
was determined from an equivalent area of sky located  circle  13\amin\, 
south in ecliptic latitude from AG Dra.

\subsubsection{Pointed Observations}

   \begin{table}
     \vspace{-0.25cm}
      \caption{Summary of \ros\, observations on AG Dra. 
             Given are for each pointing the observation ID (column 1),
             the date of the observation (2), 
             the detector (P=PSPC, H=HRI) without or with Boron (B) filter (3), 
             the nominal exposure time (4), and
             the total number of counts (5).}
            \begin{center}
            \vspace{-.2cm}
            \begin{tabular}{lrcrr}
            \hline
            \noalign{\smallskip}
            ~ROR    & Date~~~~~ & P/H & T$_{\rm Nom}$ & N$_{\rm cts}$ \\
            ~~No.    &           &     & (sec)\,       &  \\
            \noalign{\smallskip}
            \hline
            \noalign{\smallskip}
      Survey   & Nov.\,24--Dec.\,3, 1990 & P    &  2004  & 1988 \\
      200686   & April 16, 1992    & P    &  2178  & 2260 \\
      200687   & May 16, 1992      & P    &  1836  & 1761 \\
      200688   & June 13, 1992     & P    &  2210  & 2224 \\
      201041   & June 16, 1992     & P    &  1829  & 1893 \\
      201042   & Sep. 20, 1992     & P    &  1596  & 1676 \\
      201043   & Dec. 16, 1992     & P    &  1296  & 1344 \\
      200689   & March 13, 1993    & P    &  1992  & 2029 \\
      201044   & March 16, 1993    & P    &  1677  & 1761 \\
      200690   & April 15, 1993    & P    &  2450  & 2324 \\
      200691   & May 12, 1993      & P    &  2064  & 1669 \\
      180063   & Aug. 28, 1994     & H    &  1273  &   14 \\
      180063F  & Sep. 9, 1994      & P(B) &  1237  &   19 \\
      180073   & Dec. 7, 1994      & H    &  2884  &  209 \\
      180073-1 & Mar. 7, 1995      & H    &  2330  &  272 \\
      180081   & Jul. 31, 1995     & H    &  1191  &   22 \\
      180081-1 & Aug. 23, 1995     & H    &   916  &    5 \\
      180081-2 & Sep. 14, 1995     & H    &  1383  &    0 \\
      180081-3 & Feb.  6, 1996     & H    &  1551  &   88 \\
            \noalign{\smallskip}
           \hline
           \end{tabular}
           \end{center}
           \label{poin}
           \vspace{-.2cm}
   \end{table}

Several dedicated pointings on AG Dra have been performed in 1992 and 1993
with the ROSAT PSPC (Tab. \ref{poin} gives a complete log of the observations).
All pointings were performed with the target on-axis. During the last \ros\,
observation of AG Dra with the PSPC in the focal plane
(already as a TOO) the Boron filter was erroneously left in front of the 
PSPC after a scheduled calibration observation.

For each pointed observation with the PSPC in the focal plane, 
X-ray photons have been extracted within 3\amin\, of the centroid position. 
This relatively large size of the extraction circle was chosen because
the very soft photons (below channel 20) have a much larger spread in their 
measured detector coordinates.
As usual, the background was determined from a ring well
outside the source (there are no other X-ray sources within 5\amin\, of AG Dra).
Before subtraction, the background photons were normalized to the same 
area as the source extraction area.

When AG Dra was reported to go into outburst (Granslo \etal 1994) we
immediately proposed for a target of opportunity observation (TOO) with ROSAT.
AG Dra was scheduled to be observed during the last week of regular PSPC 
observations on July 7, 1994, but due to star tracker problems no photons
were collected. For all the later \ros\,
observations only the HRI could be used after the PSPC gas has
been almost completely exhausted. Consequently, no spectral information is
available for these observations. The first HRI observation took place on
August 28, 1994, about 4 weeks after the optical maximum. All the following 
HRI observations and the single PSPC observation with the Boron filter
(described above) were performed as TOO to determine the evolution of the
X-ray emission after the first outburst. With the knowledge of the results of 
the first outburst the frequency of observations was increased for the second 
optical outburst.

Source photons of the HRI observations have been extracted within 1\amin\, and 
were background and vignetting corrected in the standard manner using EXSAS 
tasks. In order to compare the HRI intensities of \ag\, with those measured 
with the PSPC we use a PSPC/HRI countrate ratio for AG Dra with its supersoft 
X-ray spectrum of 7.8 as described in Greiner \etal (1996).

\subsection{The X-ray position of AG Dra}

We derive a best-fit X-ray position from the on-axis HRI pointing 180073
of R.A. (2000.0) = 16\h01\m40\fss8,
Decl. (2000.0) = 66\grad48\amin08\asec\, with an error of $\pm$8\asec.
This position is only 2\asec\, off the optical position of AG Dra
(R.A. (2000.0) = 16\h01\m40\fss94, Decl. (2000.0) = 66\grad48\amin09\farcs7).

\subsection{The X-ray lightcurve of AG Dra}

The mean ROSAT PSPC count\-rate of AG Dra during the all-sky survey was 
determined (as described in paragraph 4.1.1) to (0.99$\pm$0.15) cts/sec.
Similar countrates were detected in several PSPC pointings during the
quiescent time interval 1991--1993.

\begin{figure*}[htbp]
      \centering{
      \hspace*{.1cm}
      \vbox{\psfig{figure=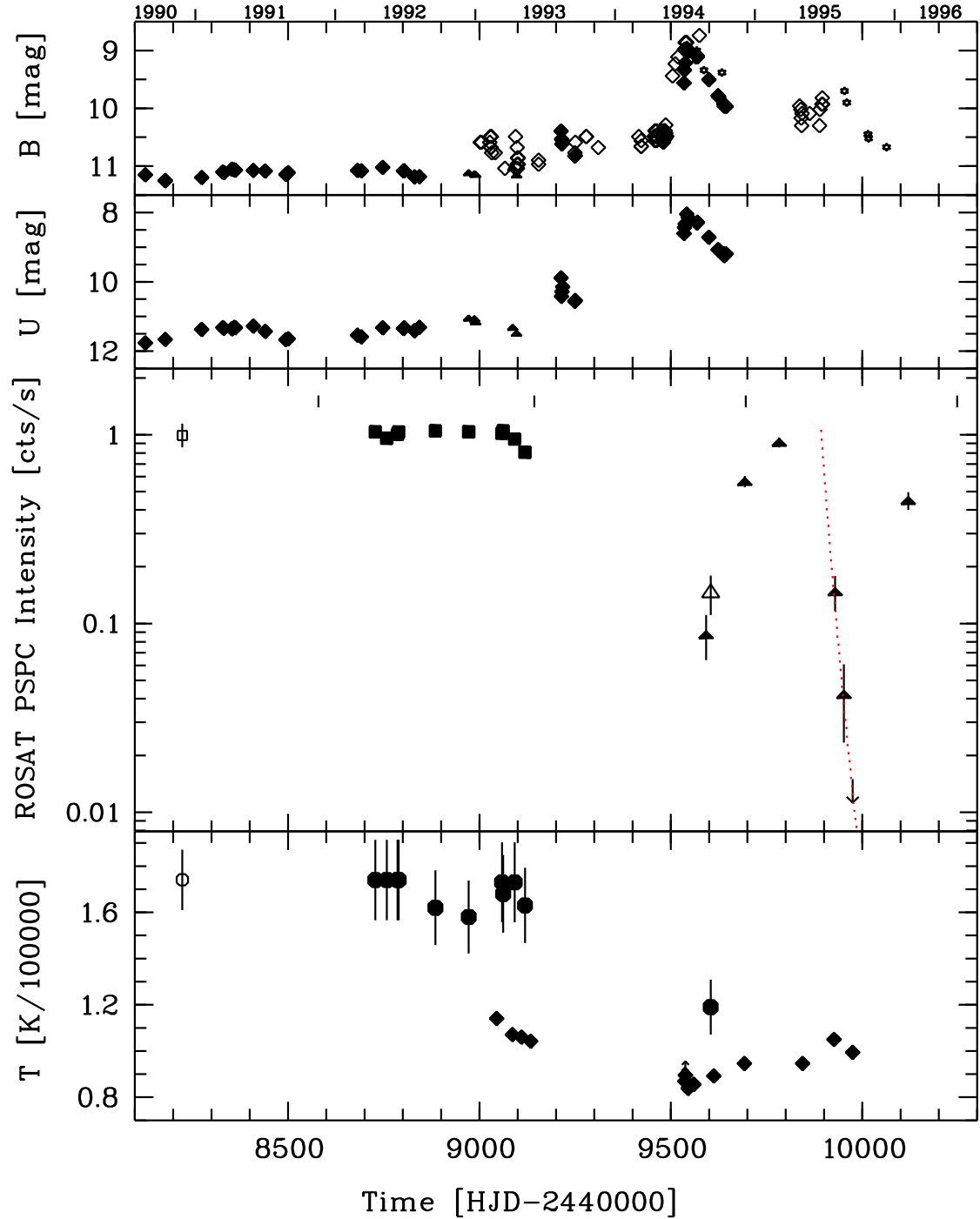,width=15cm,%
          bbllx=1.6cm,bblly=2.2cm,bburx=18.5cm,bbury=23.4cm,clip=}}\par}
      \vspace{-0.15cm}
      \caption[longlight]{The X-ray and optical light curve of AG Dra over 
           the past 5 years. The two top panels show the U and B band
           variations as observed at Sonneberg Observatory 
           (filled lozenges = photoelectric photometry, open lozenges = 
           photographic sky patrol), Skalnate Pleso Observatory
           (Hric \etal 1994; filled triangles), and italian amateurs
           (Montagni \etal 1996; stars).            The large middle panel
           shows the X-ray intensity as measured with the \ros\, satellite:
           filled squares denote PSPC observations, filled triangles
           are HRI observations with the countrate transformed to PSPC 
           rates (see text), and the open triangle is the Boron filter
           observation corrected for the filter transmission.
           Statistical errors (1$\sigma$) are overplotted; those of the 
           PSPC pointings are smaller than the symbol size.
           The vertical bars at the top indicate the minima of the
           U band lightcurve (Skopal 1994).
           The dotted line shows the fit of an expanding and cooling 
           envelope to the X-ray decay light curve.
           The lower panel shows temperature estimates from blackbody fits
           to the \ros\, X-ray data (filled circles) and from the 
           He{\II} ($\lambda$1640) flux to continuum flux at $\lambda$1340 \AA\,
              as determined from the IUE spectra and assuming E(B--V)=0.06 
              (=Zanstra temperatures, filled lozenges).}
         \label{light}
    \end{figure*}

The X-ray light curve  of \ag\, as deduced from the All-Sky-Survey data taken 
in 1990, and 11 \ros\, PSPC pointings (mean countrate over each pointing)
as well as 7 HRI pointings taken between 1991 and 1996
is shown in Fig. \ref{light}.
The countrates of the HRI pointings have been converted with a factor of 7.8
(see paragraph 4.1.2) and are also included in Fig. \ref{light}.

This 5 yrs X-ray light curve displays several features:
\begin{enumerate}
\vspace{-0.1cm}
\item The X-ray intensity has been more or less constant bet\-ween 1990 and 
the last observation (May 1993) before the optical outburst. Using the mean 
best fit blackbody model with kT = 14.5 eV and the galactic column density 
N$_{\rm H}$ = 3.15$\times$10$^{20}$ cm$^{-2}$ (fixed during the fit, 
see below), the unabsorbed intensity in the ROSAT band (0.1--2.4 keV) is 
1.2$\times$10$^{-9}$ erg cm$^{-2}$ s$^{-1}$.
\item During the times of the optical outbursts the observed X-ray flux 
drops substantially. The observed maximum amplitude of the intensity decrease 
is nearly a factor of 100.  With the lowest intensity measurement being an 
upper limit, the true amplitude is certainly even larger. Due to the poor 
sampling we can not determine whether the amplitudes of the two observed 
X-ray intensity drops are similar.
\item Between the two X-ray minima the X-ray intensity nearly reached the 
pre-outburst level, i.e. the relaxation of whatever parameter caused these drops
was nearly complete. We should, however, note that the relaxation to the 
pre-outburst level was faster at optical wavelength than at X-rays, 
and also was faster in the B band than in the U band. In 
December 1994 the optical V brightness ($\approx$9\fm5) was nearly 
back to the pre-outburst magnitude, while the X-ray intensity was still a 
factor of 2 lower than before the outburst.
\item The quiescent X-ray light curve shows two small, but significant 
intensity drops in May 1992 and April/May 1993. The latter and deeper dip
coincides with the orbital minimum in the U lightcurve (Meinunger 1979,
Skopal 1994) and might suggest a periodic, orbital flux variation.
\end{enumerate}

   \begin{figure*}[htbp]
      \centering{
      \hspace*{.1cm}
      \vbox{\psfig{figure=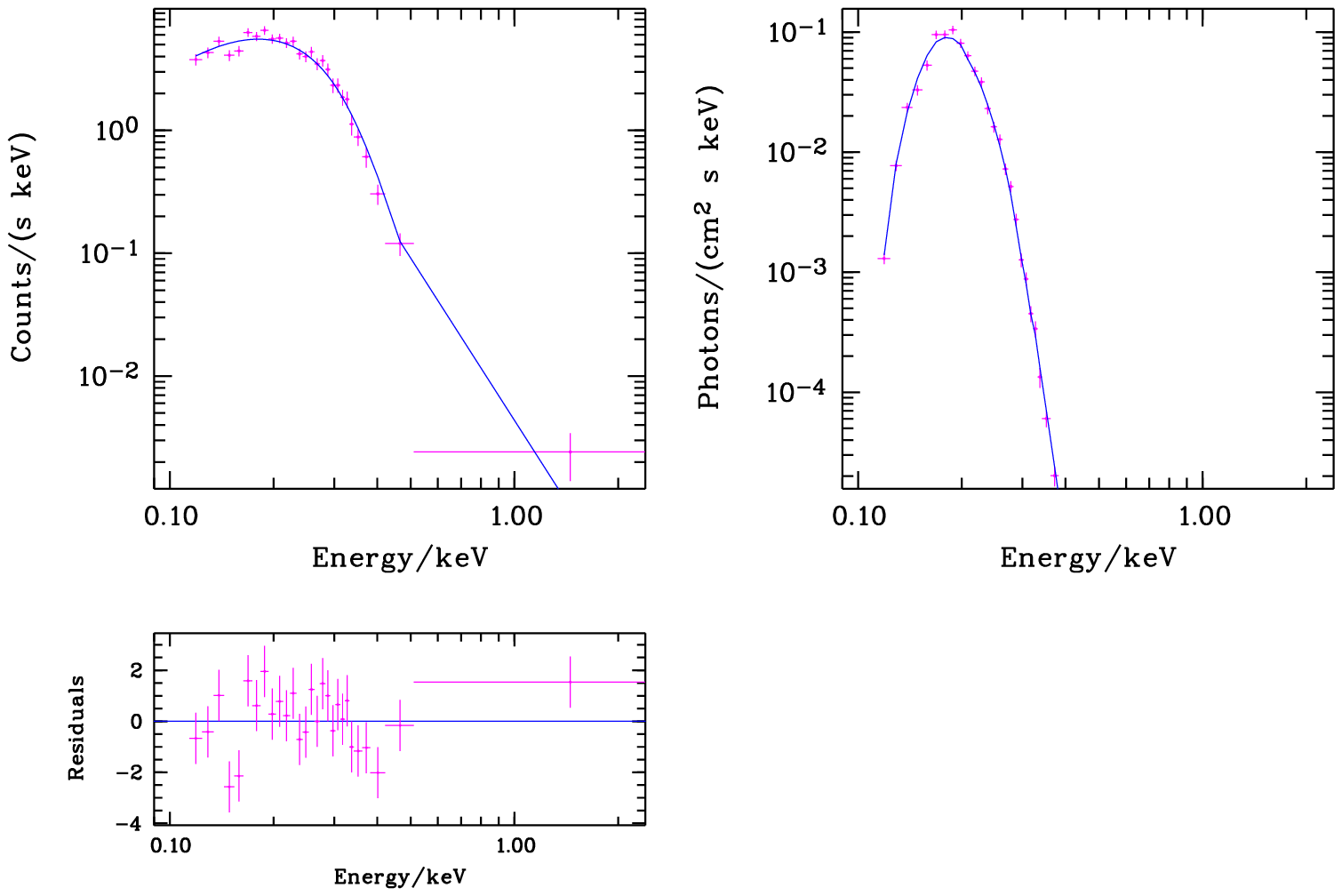,width=15cm,%
          bbllx=2.5cm,bblly=1.3cm,bburx=17.8cm,bbury=11.5cm,clip=}}\par
      \vspace{-3.2cm}\hspace{8.5cm}
      \vbox{\psfig{figure=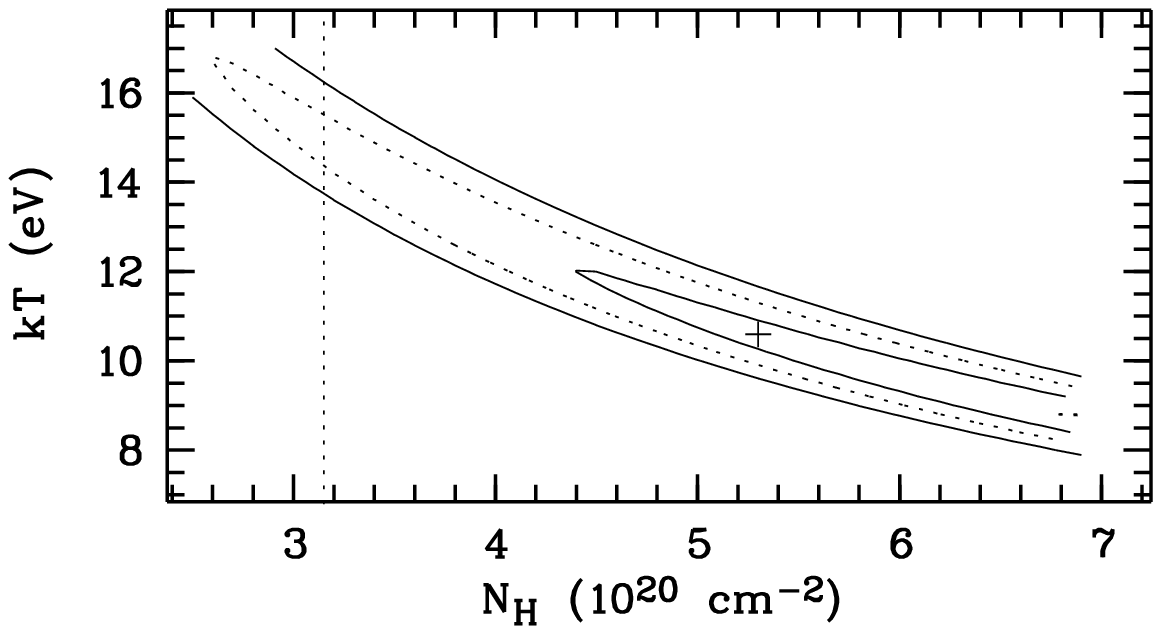,width=6.5cm,%
          bbllx=2.1cm,bblly=20.8cm,bburx=13.9cm,bbury=27.2cm,clip=}}\par}
      \caption[xspectrum]{ROSAT X-ray spectrum of AG Dra in quiescence
                on April 15, 1993. A blackbody model is used; for best fit 
                parameters see Tab. \ref{xfitpar}. 
                The upper left panel shows the countrate spectrum, while
                the upper right panel shows the absorbed photon spectrum
                after folding with the detector response.
                The lower left panel shows the residuals of the fit in units 
                of $\sigma$, while the lower right panel
                visualizes the range of 1, 2 and 3 $\sigma$ contours of 
                the absorbing column and the best fit blackbody temperature
                in the three parameter fit. The cross marks the best fit
                and the dotted line marks the total galactic column density.}
         \label{xspec}
    \end{figure*}

\subsection{The X-ray spectrum of AG Dra in quiescence}

For spectral fitting of the all-sky-survey data 
the photons in the amplitude channels 11--240 
(though there are almost no photons above channel 50) were binned 
with a constant signal/noise ratio of 9$\sigma$. The fit of a blackbody
model with all parameters left free results in an effective temperature of 
kT$_{\rm bb}$ = 11 eV (see Tab. \ref{xfitpar}).

Since the number of counts detected during the individual PSPC pointings 
allows high signal-to-noise spectra, we investigated the possibility of 
X-ray spectral changes
with time. First, we kept the absorbing column fixed at its galactic
value and determined the temperature being the only fit parameter. We find
no systematic trend of a temperature decrease (lower panel of Fig. \ref{light}).
Second, we kept the temperature fixed (at 15 eV in the first run and at the
best fit value of the two parameter fit in the second run) and checked 
for changes in N$_{\rm H}$, again finding no correlation.
Thus, no variations of the X-ray spectrum could be found along the orbit.
The rather small degree of observed variation of temperature and N$_{\rm H}$ 
(Tab. \ref{xfitpar}) over more than three years during quiescence 
(including the \ros\, All-Sky-Survey data) are not regarded to be significant
due to the correlation of these quantities (Fig. \ref{xspec}).

The independent estimate of the absorbing column towards AG Dra from
the X-ray spectral fitting indicates that the detected AG Dra emission 
experiences the full galactic absorption. While fits with N$_{\rm H}$ as free
parameter (see Tab. \ref{xfitpar}) systematically give values slightly higher 
than the galactic
value (which might led to speculations of intrinsic absorption), we assess
the difference to be not significant due to the strong interrelation
of the fit parameters (see lower right panel of Fig. \ref{xspec}) 
given the energy resolution of the PSPC and the softness
of the X-ray spectrum. We will therefore use the galactic N$_{\rm H}$ value
(3.15$\times$10$^{20}$ cm$^{-2}$ according to Dickey and Lockman 1990) 
in the following discussion.

   \begin{table}
     \vspace{-0.15cm}
      \caption{Summary of blackbody model fits to the \ros\, PSPC observations 
             of AG Dra during quiescence. 
             Fluxes are in photons/cm$^2$/s and temperatures kT in eV.
             The absorbing column N$_{\rm H}$ is in units of 10$^{20}$ cm$^{-2}$
             for the three parameter fit and was fixed at the galactic value
             of 3.15$\times$10$^{20}$ cm$^{-2}$ for the two parameter fit.}
            \begin{center}
            \vspace{-.2cm}
            \begin{tabular}{ccccccc}
            \hline
            \noalign{\smallskip}
            ROR  & Date & \multicolumn{3}{c}{3-parameter fit} & 
                            \multicolumn{2}{c}{2-parameter fit} \\
             No. & &  N$_{\rm H}$ & Flux & kT    &  ~~~Flux~~~ & kT  \\
            \noalign{\smallskip}
            \hline
            \noalign{\smallskip}
   Survey  & 901129 & 5.0 &  7040 &  10.6      &   172.5  &  14.6 \\
   200686  & 920416 & 5.1 &  4680 &  11.0      &   144.2  &  15.0 \\
   200687  & 920516 & 4.6 &  1280 &  11.7      &   147.0  &  15.0 \\
   200688  & 920613 & 4.9 &  2750 &  11.2      &   145.0  &  15.0 \\
   201041  & 920616 & 4.7 &  1520 &  11.6      &   146.7  &  15.0 \\
   201042  & 920920 & 3.5 &  \,\,860 &  13.2      &   304.1  &  14.0 \\
   201043  & 921216 & 4.2 &  1360 &  11.3      &   421.1  &  13.6 \\
   200689  & 930313 & 5.2 &  9060 &  10.6      &   160.1  &  14.9 \\
   201044  & 930316 & 4.5 &  1240 &  11.6      &   205.8  &  14.5 \\
   200690  & 930415 & 5.3 &  9760 &  10.6      &   147.7  &  14.9 \\
   200691  & 930512 & 5.1 &  9750 &  10.4      &   227.9  &  14.1 \\
            \noalign{\smallskip}
           \hline
           \end{tabular}
           \end{center}
           \label{xfitpar}
           \vspace{-.2cm}
   \end{table}

With N$_{\rm H}$ fixed at its galactic value the mean temperature during 
quiescence is about 
14--15 eV, corresponding to 160000--175000 K. These best fit temperatures
are plotted in a separate panel below the X-ray intensity (Fig. \ref{light}).
The small variations in temperature during the quiescent phase are consistent
with a constant temperature of the hot component of AG Dra.

\subsection{The X-ray spectrum of AG Dra in outburst}

As noted already earlier (e.g. Friedjung 1988),
the observed fading of the X-ray emission during the optical outbursts
of AG Dra can be caused either by a temperature decrease of the hot
component or an increased absorbing layer between the X-ray source and
the observer. In order to evaluate the effect of these possibilities, 
we have performed model calculations using the response of the \ros\, HRI.
In a first step, we assume a 15 eV blackbody model and determine the increase
of the absorbing column density necessary to reduce the \ros\, HRI countrate by 
a factor of hundred. The result is a factor of three increase.
In a second step we start from the two parameter best fit and determine
the temperature decrease which is necessary to reduce the \ros\, HRI countrate
at a constant absorbing column (3.15$\times$10$^{20}$ cm$^{-2}$). 
We find that the temperature of the hot component has to decrease from
15 to 10 eV, or correspondingly from 175000 K to 115000 K.

The only ROSAT PSPC observation (i.e. with spectral resolution) 
during optical outburst is the one with Boron filter. 
The three parameter fit as well as the two parameter
fit give a consistently lower temperature. But since the Boron filter cuts
away the high-end of the Wien tail of the blackbody, and we have only 19 
photons to apply our model to, we do not regard this single measurement
as evidence for a temperature decrease during the optical outburst.

What seems to be excluded, however, is any enhanced absorbing column during
the Boron filter observation. The best fit absorbing column of the three 
parameter fit is 4.4$\times$10$^{20}$ cm$^{-2}$, consistent with the
best fit absorbing column during quiescence. Since the low energy part of the
spectrum in the PSPC is not affected by the Boron filter except a general
reduction in efficiency by roughly a factor of 5, any increase of the absorbing
column would still be easily detectable. For instance, an increase of the
absorbing column by a factor of two (to 6.3$\times$10$^{20}$ cm$^{-2}$)
would absorb all photons below 0.2 keV
and would drop the countrate by a factor of 50 contrary to what is observed.

It is interesting to note that the decrease of the X-ray flux is similarly 
strong in both, the 1994 and 1995 outbursts, while the optical amplitude of 
the secondary outburst in 1995 was considerably smaller than the first 
outburst. We note in passing that the intensity of the He{\II} and N{\V} lines
also showed a comparable large increase in the main 1981/1982 outburst and the
minor outburst in 1985/1986 
while the optical and the short wavelength UV continuum 
amplitudes again were smaller in the latter outburst (Mikolajewska \etal 1995).
Since the short wavelength UV continuum is the Rayleigh-Jeans tail of the hot 
(blackbody) component, this behaviour suggests that a temperature decrease is 
the cause of the reduced X-ray intensity during outburst rather than increased 
absorption with the temperature decrease being smaller in the secondary 
outbursts.

\section{Discussion}

\subsection{Quiescent X-ray emission}

\subsubsection{The X-ray spectrum and luminosity}

Previous analyses of X-ray emission of symbiotic stars have been interpreted 
either with blackbody (Kenyon and Webbink 1984) or NLTE atmosphere 
(Jordan \etal 1994) models representing the hot component,
or with bremsstrahlung emission from a hot, gaseous nebula 
(Kwok and Leahy 1984). Our ROSAT PSPC spectra of AG Dra show no hint for any 
hard X-ray emission. The soft spectrum is well fitted with a blackbody model.
A thermal bremsstrahlung fit is not acceptable to this soft energy distribution.
There is also no need for a second component as proposed by Anderson \etal 
(1981): a fit of a blackbody and
thermal bremsstrahlung model with the bremsstrahlung temperature limited
to values above 0.1 keV results in an unabsorbed flux ratio 
between blackbody and bremsstrahlung of 85000:1 in the 0.1--2.4 keV band.
A hard X-ray component could be expected from an accretion disk
both for radial and disk accretion (Livio 1988) at the relatively 
low accretion rate implied for this system. This absence of any hard
emission also rules out the possibility of interpreting the soft spectrum
as arising in the boundary layer. The most extreme ratios for soft to hard
emission observed sofar in magnetic cataclysmic variables are of the order
of 1000:1. An example of the blackbody fit is given in 
Fig. \ref{xspec} using the observation of April 1993 (having the largest 
number of photons). 

The simple model of explaining the soft X-ray emission by an accretion disk 
is ruled out not only due to the soft spectrum, but also due to the observed 
luminosity which is much higher (see below) than can be produced by an 
accretion disk around a compact dwarf and the extremely low accretion rate.
If the X-ray spectrum were to explained by a standard accretion disk model
(Shakura \& Sunyaev 1973) then the necessary accretion rate would be
of the order of 1.7$\times$10$^{-7}$ {$\dot M$}$_{\rm edd}$ (or
4.5$\times$10$^{-15}$ \msun/yr) for a 1 \msun\, compact object. This is 
unreasonably small, basically due to the low effective temperature
(T$_{\rm max}^{\rm eff}$ = 24 eV) of the
X-ray emission coupled with a small mass compact object. 

\begin{figure}[htbp]
      \centering{
      \vbox{\psfig{figure=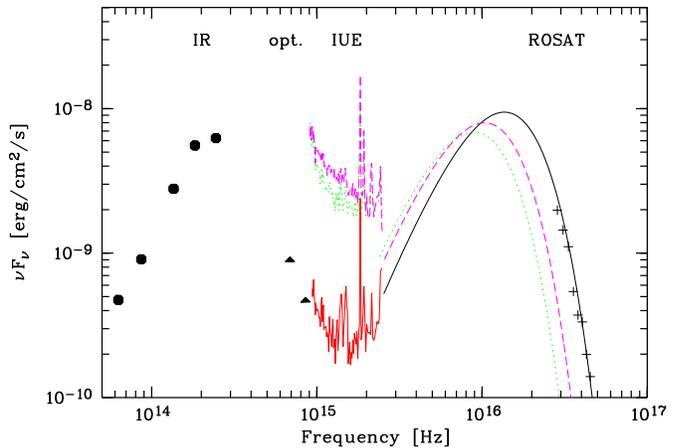,width=8.8cm,%
          bbllx=1.4cm,bblly=15.1cm,bburx=19.cm,bbury=27.2cm,clip=}}\par}
      \vspace{-0.15cm}
      \caption[nunfnu]{Energy distribution of AG Dra during quiescence
          and outburst: The full line shows the quiescent spectrum,
          combined from 
          a UV spectrum of April 9, 1993, and a \ros\, observation
          of April 15, 1993 (best fit blackbody extrapolation is shown
          on top of the absorption corrected data points). The optical
          B measurement is also from April 15, while the U measurement
          is from April 10, 1993. The dashed/dotted lines are the
          measured UV spectra during outburst of July 28 / Sep 14, 1995
          together with the modelled blackbody spectra of lower temperature
          (11/9.5 eV) which reproduce the observed HRI countrates as 
          measured with \ros\, on July 31 / Sep. 14, 1995.
              }
         \label{qnufnu}
    \end{figure}

Using the blackbody fit parameters while N$_{\rm H}$ was fixed at its 
galactic value, i.e. kT = 14.5 eV and a normalization parameter of 206
photons/cm$^2$/s corresponding to observation 201044 
(from experience with soft ROSAT spectra we know that this
procedure helps to avoid overpredicting the flux using blackbody models), 
the unabsorbed bolometric 
luminosity of the hot component in AG Dra during quiescence (1990--1993) is 
(9.5$\pm$1.5)$\times$10$^{36}$ (D/2.5 kpc)$^2$ erg/s 
(or equivalently 2500$\pm$400 (D/2.5 kpc)$^2$ \lsun)
with an uncertainty of a factor of a few due to the errors in the 
absorbing column and the temperature (see Fig. \ref{qnufnu} for a visualization
of the fact that the ROSAT measurement actually covers only the tail
of the spectral energy distribution). The blackbody radius is derived
to be R$_{\rm bb}$ = (4.1$\pm$1.5)$\times$10$^9$ cm (D/2.5 kpc). 
(Previous estimates arrived at much lower values due to the overestimate of 
the temperature.) 

This high luminosity during quiescence and the size of 
the emitting region comparable to a white dwarf radius suggests
that the primary is a white dwarf in the state of surface hydrogen burning.
Assuming that the bolometric luminosity of the hot component equals the 
nuclear burning luminosity, the burning rate is
\mdot $\approx$ (3.2$\pm$0.5)$\times$10$^{-8}$ (D/2.5 kpc)$^2$ \msun/yr.
Under the assumption of a steady state the same amount of matter
should be accreted onto the white dwarf from the companion 
(via wind or an accretion disk, see below).
Indeed, this burning rate lies within the stable-burning regime for a white 
dwarf of less than 0.6 \msun\, (Iben and Tutukov 1989) consistent with the
core mass-luminosity relation L/\lsun $\approx$ 4.6$\times$10$^4$ 
(M$_{\rm core}$/\msun -- 0.26) (Yungelson \etal 1996).
From the orbital parameters (Mikolajewska \etal 1995, Smith \etal 1996) and an 
inclination less than about 85\grad\, (see paragraph 5.1.2) the donor mass is 
estimated to be lower than 2 \msun, in agreement with the estimated surface 
gravity of Smith \etal (1996).

\subsubsection{Orbital flux modulation?}

The observational coverage of AG Dra over 1992/1993 is extensive enough to 
also investigate possible temporal and spectral changes of the X-ray flux 
along the orbital phase. Using the ephemeris from Skopal (1994) the minimum 
phases of AG Dra occurred in November 1991 and  June 1993. 

The coincidence of the U minimum and the drop in X-ray flux in mid-1993 is 
surprising, suggesting the possible discovery of orbital modulation of the 
X-ray flux. Unfortunately, we have no full coverage of the orbital period, and 
thus the duration of the flux depression can not be determined. But even with 
the available data this duration is a matter of concern: The X-ray intensity
decrease occurs
very slowly over a time interval of two months, suggesting a total duration
of at least 4 months if this modulation is symmetric in shape.

The radial velocity data demonstrate that during the U minima the cool 
companion lies in front of the hot component (cf. Mikolajewska \etal 1995).
Depending on the luminosity class, the maximum duration for an eclipse of
the WD by the giant (bright giant) companion is 10 (24) days, i.e. much shorter
than the observed time scale at X-rays.
Thus, the drop in X-ray intensity cannot be due to a WD eclipse.
As evidenced by near-simultaneous IUE spectra between April and June 1993,
the far-UV continuum associated to the tail of the hot component's
continuum emission is not occulted, supporting the previous argument.
A similar conclusion was reached when considering the nearly complete lack of
orbital variations of the short wavelength UV continuum (Mikolajewska \etal 
1995). By inverting the arguments, the lack of an X-ray eclipse implies
that the inclination should be less than 87\grad (82\grad) for a giant
(bright giant) with 20 \rsun\, (70 \rsun) radius.

Even in the reflection model of Formiggini \& Leibowitz (1990), 
in which the eclipse of the hot component is short
and its depth is expected to be considerably larger in the UV (and possibly
in the soft X-ray band) than in the optical, it is difficult to imagine
that a possible UV eclipse has been gone undetected.

Alternatively to an eclipse interpretation of the X-ray intensity drop
just before the 1993 U band minimum, one could think of a pre-outburst
which was missed in the optical region except for the marginal flux increase 
in the B band after JD = 244\,9000. In this case the interpretation
would be similar to that of the major X-ray intensity drops during the optical
outbursts in 1994 and 1995 mass loss (see below). Independent of the actual 
behaviour around
JD = 9100 it is interesting to note that there is a secure detection of
a 2 mag U brightness jump coincident with a nearly 1 mag B brightness
increase shortly after JD = 9200. A slight brightening by about 0.2--0.3 mag
is also present in the AAVSO light curve around JD 9205--9215 (Mattei 1995).

\subsubsection{Wind mass loss of the donor and accretion}

Accepting the high luminosity during quiescence and consequently assuming that 
the hot component in the AG Dra system is in (or near to) the surface hydrogen
burning regime, the companion has to supply the matter at the high rate 
of consumption by the hot component.

It is generally assumed (and supported by three-dimensional gas dynamical 
calculations of the accretion from an inhomogeneous medium) that the 
compact objects in symbiotic systems accrete matter from the wind of 
the companion according to the classical Bondi-Hoyle formula.
The wind mass loss rate of the companion depends on several parameters. 
One of the important ones is the luminosity which governs the location with 
respect to the Linsky and Haisch dividing line. Accordingly, a bright giant 
(luminosity class II) is expected to have a considerably larger mass loss rate 
than a giant (luminosity class III).

As mentioned already earlier, the spectral classification and the luminosity
class of the companion in the AG Dra binary system is not yet securely 
determined. According to the formula of Reimers (1975):
  $$ - \dot M = 4 \times 10^{-13} \eta {R L \over M} M_{\odot}/yr $$
where R, L and M are the radius, luminosity and mass of the star in solar
units, and $\eta$ a constant between 0.1 and 1, depending primarily on the
initial stellar mass, the commonly used K3III classification would imply
a mass loss rate of (0.2--7)$\times$10$^{-10}$ \msun/yr, i.e. a factor of 
100 lower than the rate necessary for steady state burning.

However, there are two additional observational hints which might help resolve 
the discrepancy between the expected mass loss of the cool donor and the
necessary rate for a steady state burning white dwarf:
(1) The cool component of AG Dra might be brighter than an average 
solar-metallicity giant. A comparison of the luminosity functions of
K giants with 4000 \lax \,T$_{\rm eff}$ \lax \,4400 in a low-metallicity versus
solar-metallicity population has shown convincingly that low-metallicity
K giants are nearly 2 mag brighter than the solar-metallicity giants
(Smith \etal 1996).
(2) Low-metallicity giants might have larger mass-loss rates than
usual giants. From observed larger 12$\mu$m excess in symbiotics as compared 
to that in normal giants a larger mass loss of giants in symbiotics has
been deduced (Kenyon 1988). The recent comparison of IR excesses of 
d-type symbiotics with that of CH and barium stars supports this evidence
(Smith \etal 1996).
Both of these observational hints argue for a mass loss of the cool component
in AG Dra which is higher than the value derived from Reimer's formula.
At present it seems premature, however, to conclude that the wind mass loss
of the donor in AG Dra is large enough to supply the matter for a steady
state surface burning on the white dwarf.

An additional problem with spherical accretion at very high rates is the
earlier recognized fact (Nussbaumer \& Vogel 1987) that the density of matter 
in the vicinity of the accretor is n$_{\rm H} \approx$ 2$\times$10$^7$ 
(\mdot/10$^{-7}$ \msun/yr)\,(V/10 km/s)$^3$\,(M$_{\rm WD}$/\msun)$^{-2}$ 
cm$^{-3}$ (Yungelson \etal 1996), i.e. the soft X-rays of the burning
accretor will be heavily absorbed.

Alternatively, one might consider the possibility that the donor
overfills its Roche lobe and that the matter supply to the white dwarf occurs 
via an accretion disk. Several issues have to be considered:
(1) Roche-lobe filling: In order to fill its Roche lobe, the donor has to
be either rather massive (as compared to a usual K giant) or rather luminous. 
However, a donor mass larger than $\approx$2\,\msun\, seems difficult due to 
the high galactic latitude and proper motion as well as on evolutionary 
grounds. Similarly, a binary system with a bright giant implies a larger
distance as compared to a giant companion.
(2) Evidence for the existence of an accretion disk:
Indeed, Garcia (1986) has raised the suspicion that AG Dra 
contains an accretion disk. Circumstantial evidence for this comes from 
predictions of Roche lobe overflow, rapid flickering on timescales of 
the order of minutes in the optical band and the observation of double-peaked 
Balmer emission lines.
Robinson \etal (1994) have investigated in detail the double-peaked
line profile of the Balmer lines in several symbiotic stars.
In the case of AG Dra they find double-peaked emission lines only in one out 
of three observations which were three and one year apart, respectively. 
Using an inclination of i=32\grad\, as proposed by Garcia (1986)
the double-peaked profile gives an acceptable fit for an 
accretion disk with an inner radius of 1.1$\times$10$^8$ cm and an outer
radius of 1.3$\times$10$^{10}$ cm though the asymmetry may be explained also
by self-absorption (Tomov \& Tomova 1996).
Recent high-resolution spectroscopy of AG Dra in the red wavelength region 
using AURELIE at OHP
performed in December 1990 and January 1995 (i.e. during very different
phases in the AG Dra orbit) revealed only a single peaked H$\alpha$
profile (Rossi \etal 1996) and Dobrzycka \etal (1996) failed to find
evidence for flickering. Both of these new observational results argue 
against a steady accretion disk in the AG Dra system.

\subsection{Quiescent UV emission}

UV observations show that the hot components of symbiotic stars are located
in the same quarters of the Hertzsprung-Russell diagram as the central stars of
planetary nebula (M\"{u}rset \etal 1991). Due to the large binary separation
in symbiotic systems the present hot component (or evolved component)
should have evolved nearly undisturbed through
the red giant phase. However, the outermost layers of the white dwarf
might be enriched in hydrogen rich material accreted from the cool companion.

Presently the far-UV radiation of the WD is ionizing a circumstellar
nebula, mostly formed (or filled in) by the cool star wind.
The UV spectroscopy suggests that the CNO composition of the nebula
is nitrogen enriched (C/N=0.63, and (C+N)/O=0.43), which is typical of
the composition of a metal poor giant atmosphere after CN cycle burning
and a first dredge up phase (Schmidt \& Nussbaumer 1993). 
We recall that recently a low metal abundance of the K-star photosphere
of [Fe/H]=--1.3 was derived (Smith \etal 1996) in agreement with it being 
a halo object.
The UV spectrum in quiescence is the usual in symbiotic systems: a continuum
increasing toward the shortest wavelengths with strong narrow emission lines
superimposed. The most prominent lines in the quiescence UV spectrum of
AG Dra are, in order of decreasing intensity: 
He{\II} 1640 \AA, C{\IV} 1550 \AA,
N{\V} 1240 \AA, the blend of Si{\IV} and O\IV] at 1400 \AA, N\IV] 1486 \AA\
and  O\III] 1663 \AA. The continuum becomes flatter longward approximately
2600 \AA\ due to the contribution of the recombination continuum originated
in the nebula. Although both continuum and emission lines have been found to
be variable during quiescence (e.g. Mikolajewska \etal 1995), there is no
clear relation with the orbital period of the system. 
The ratio intensity of the recombination He{\II} 1640\AA\, line
to the far-UV continuum at 1340 \AA\, suggests a Zanstra temperature of
around 1.0-1.1$\times$10$^5$ K during quiescence (see Fig. \ref{light}).

The large X-ray luminosity together with its soft spectrum allows to understand 
the large flux from He{\II}, most notably $\lambda$4686 \AA\, and 
$\lambda$1640 \AA, at quiescence (which could not be explained in earlier 
models; see e.g. Kenyon \& Webbink 1984) as being due to X-ray ionization.

\subsection{X-ray emission during the optical outbursts}

\subsubsection{Previously proposed models}

Three different basic mechanisms have been proposed to explain the drastic 
intensity changes of symbiotic stars during the outburst events: 
(1) Thermonuclear runaways on the surface of the hot component (WD) after 
the accreted envelope has reached a critical mass (Tutukov \& Yungelson 1976,
Paczynski \& Rudak 1980). The characteristic features are considerable 
changes in effective temperature at constant bolometric luminosity, and the
appearance of an A--F supergiant spectrum thought to be produced by the 
expanding WD shell.
(2) Instabilities in an accretion disk after an increase of mass transfer
from the companion (Bath \& Pringle 1982, Duschl 1986). 
(3) Ionization changes of the H{\II} region (from density-bounded to 
radiation-bounded) around the hot component caused by an abrupt change in 
the mass loss rate of the companion (Nussbaumer \& Vogel 1987, Mikolajewska \& 
Kenyon 1992). 
In the case of AG Dra, all these three scenarios  have problems with some 
observational facts. The thermonuclear runaway is rejected by the fact that
the quiescent luminosity is already at a level which strongly suggests
burning before the outbursts. The disk instability scenario predicts
substantial variation of the bolometric luminosity during the outbursts,
for which no hints are available in AG Dra. The application of ionization
changes to AG Dra is questionable because the nebula is apparently not in 
ionization equilibrium (Leibowitz \& Formiggini 1992).

Specifically for AG Dra an additional scenario has been proposed, namely 
(4) the liberation of mechanical energy in the atmosphere of the companion 
(Leibowitz and Formiggini 1992).

\subsubsection{Expanding and contracting white dwarf}

Using our finding of an anticorrelation of the optical and X-ray intensity
and the lack of considerable changes in the temperature of the hot component
during the 1994/95 outburst of AG Dra, we propose the following rough scenario.
(1) The white dwarf is already burning hydrogen stably on its surface 
    before the optical outburst(s).
(2) Increased mass transfer, possibly episodic, from the cool companion 
    results in a slow expansion of the white dwarf.
(3) The expansion is restricted either due to the finite excess mass accreted 
    or by the wind driven mass loss from the expanding photosphere of the 
    accretor. This wind possibly also suppresses further accretion onto the 
    white dwarf. The photosphere is expected to get cooler with increasing 
    radius.
(4) The white dwarf is contracting back to its original state once the
    accretion rate drops to its pre-outburst level.
    Since the white dwarf is very sensitive to its boundary conditions,
    it is not expected to return into a steady state immediately
    (Paczynski \& Rudak 1980). Instead, it
    might oscillate around the equilibrium state giving rise to secondary
    or even a sequence of smaller outbursts following the first one.

The scenario of an expanding white dwarf photosphere due to the increase in 
mass of the hydrogen-rich envelope has been proposed already by 
Sugimoto \etal (1979) on theoretical grounds without application to a specific
source class. The expansion velocity was shown to be rather low (Fujimoto 1982)

$$
 {dR \over dt} = R {d \ln R \over d \ln \Delta M_1} \, 
     \left ({ {\dot M - \dot M_{\rm RG}} \over \Delta M_1}\right ) \qquad $$
$$  \qquad\qquad\qquad  \approx {8~} 
  \left ({\dot M - \dot M_{\rm RG}} \over {10^{-6} M_{\odot} yr^{-1}}\right ) 
  \left (\Delta M_1 \over {10^{-5} M_{\odot}}\right )^{-1} m s^{-1}
$$

where $d \ln R / d \ln \Delta M_1 \approx 3-4$ at $R \approx 1 R_{\odot}$ was
adopted, appropriate for a low mass white dwarf accreting at high rates
(Fujimoto 1982). We note in passing that this value could be different at 
smaller  radii, i.e. near white dwarf dimensions (see e.g. Neo \etal (1977)). 
In particular, depending on the location in the H-R diagram it would be 
around 2.3 on the high-luminosity plateau of a 0.6 $M_{\odot}$ white dwarf, 
but rising up to 7 around the high-temperature knee (Bl\"ocker 1996), thus
suggesting  a non-linear expansion rate.
If we assume that the luminosity remains constant during the expansion
we can determine the expansion velocity 
simply by folding the corresponding temperature 
decrease by the response of the ROSAT detector.
Fitting the countrate decrease of a factor of 3.5 within 23 days (from
0.14 cts/s on HJD 9929 to 0.04 cts/s on HJD 9952 corresponding to a 
concordant temperature decrease from 12.3 eV to 11.1 eV) we find that
$$\left (\dot M - \dot M_{\rm RG} \over 10^{-6} M_{\odot} yr^{-1}\right ) 
\left (\Delta M_1 \over {10^{-5} M_{\odot}}\right )^{-1} \approx 0.8 $$
for the input parameters (blackbody radius) corresponding to 2.5 kpc
(note that the input parameters change with distance).
The mass of the envelope $\Delta M_1$ is difficult to assess due to the
non-stationarity in the AG Dra system. In thermal equilibrium at our
derived mean temperature it should be larger than \gax5$\times$10$^{-5}$ \msun\,
for a white dwarf with mass M $<$ 0.6 \msun\, in the steady
hydrogen-burning regime (Fujimoto 1982, Iben \& Tutukov 1989).
Inserting this in the above equation implies that the accretion rate
necessary to trigger an expansion which is consistent with the X-ray 
measurements has to be (8.1/2.5/1.4)$\times$10$^{-6}$ \msun/yr for
white dwarf masses of 0.4/0.5/0.6 \msun. This is a factor of 40--250 larger 
than the quiescent burning/accretion rate.
The corresponding mean expansion rate is $dR/dt$ $\approx$6.5 m/s. 
With a duration of the X-ray decline in 1995 of about 100 days (which is
consistent with the 1994 rise time in the optical region) we find that
during the 1995 outburst (and probably also during the 1994 outburst)
the radius of the accretor roughly doubled while the temperature decreased
by about 35\% (from 14.5 eV to 9.5 eV for the two parameter fit).
The countrate decrease modelled with the above parameters is shown as the 
dotted line in Fig. \ref{light},
and the extrapolation to the quiescent countrate level gives an expected
onset of the expansion around HJD = 9902.

A completely different and independent estimate of the response time of
a white dwarf gives a similar result, thus  
it seems quite reasonable that a white dwarf can indeed expand and contract 
on a timescale (see e.g. Kovetz \& Prialnik 1994, Kato 1996) which
is observed as optical outburst rise and fall time in AG Dra.
The contraction timescale can be approximated by the duration of the
mass-ejection phase (Livio 1992), similar to the application to the
supersoft transient RX J0513.9--6951 (Southwell \etal 1996),
where $\tau$\,$\approx$\,51\,/\,\mt\,(\mt$^{-2/3}$ -- \mt$^{2/3}$)$^{3/2}$ days
with \mt\, being the ratio of white dwarf mass to the Chandrasekhar mass
(Livio 1992). This relation gives a timescale of the order of 100 days
(as observed) for a white dwarf mass of 0.5--0.6 \msun, consistent with
observations of the AG Dra optical outbursts.

Given the substantial accretion rate triggering the optical outbursts
(which has to be supplied by the donor), 
mass loss via a wind from the cool companion
seems to be too low to power the outbursts. Depending on the donor state
in quiescence (where wind accretion is possible though we prefer Roche
lobe filling; see paragraph 5.1.3.), two possibilities are conceivable: Either
the companion fills its Roche lobe all the time, and the outbursts (i.e. the
increased mass transfer) are triggered by fluctuations of the cool
companion (e.g. radius), or a more massive and/or more luminous 
companion produces a strong enough wind to sustain the burning
in the quiescent state without filling its Roche lobe and only
occasionally overfills its Roche lobe thus triggering the outbursts.
As noted above, a Roche lobe filling giant implies a distance larger than 
the adopted 2.5 kpc. While this imposes no problems with the interpretation
of our UV and X-ray data (in fact a larger distance implies a larger
intrinsic luminosity and thus shifts AG Dra even further into the stability 
burning region for even higher white dwarf masses), the distance dependent 
numbers derived here have to be adapted accordingly.

\subsubsection{Wind from the accretor}

If the accretion indeed is spherical (i.e. as wind from the donor), it may
occasionally be suppressed by the wind from the accretor (Inaguchi \etal 1986).
Hot stars with radiative envelopes are thought to suffer intense 
(radiation-driven) stellar winds at a rate of 
\mdot$_{\rm wind loss}$ = L/(v$_{\rm esc}$ c) where L is the burning
luminosity, v$_{\rm esc}$ is the escape velocity and c the speed of light.
This wind becomes increasingly effective at larger radii. Thus, a
radiatively driven wind from the WD (Prialnik, Shara \& Shaviv 1978,
Kato 1983a,b) might restrict the expansion of a RG-like envelope or 
an expelled shell or even might remove the envelope (Yungelson \etal 1995).
An optically thick wind can be even more powerful by up to a factor of 10
(Kato and Hachisu 1994).

The wind of the hot white dwarf has typical velocities of a few hundreds
km/s (Kato \& Hachisu 1994). At these velocities it would take only
several days until this white dwarf wind reaches the Roche lobe of the
cool component, i.e. the wind of the cool component. This timescale
is short enough to possibly cause the variations in the intensity of those
lines which are thought to arise at the illuminated side of the wind zone 
of the cool component.

\subsubsection{Alternative scenario}

An alternative to increased mass transfer would be to invoke a mild He flash 
on the hydrogen burning white dwarf which again
might cause the photosphere to expand. 
Previous investigations for mild flashes have yielded timescales of the order
of 15--20 yrs even for the most massive white dwarfs (Paczynski 1975).
Recent calculations of H and He flashes
have shown that under certain circumstances  flash ignition can be rather
mild without leading to drastic explosive phenomena, and shorter than
about 1 yr already for white dwarf masses below 1 \msun\, (Kato 1996).

\subsection{UV emission during the optical outbursts}

There is an evident mismatch between the outburst UV spectra and the
extrapolation toward shorter wavelengths of the blackbodies with
temperatures inferred from the ROSAT data (Fig. \ref{qnufnu}). 
The most plausible explanation
is the existence of an additional emission mechanism in the UV, namely
recombination continuum in the nebula surrounding the hot star. While during
the quiescent phase the contribution of the nebular continuum is not very
large (although it is not negligible), it increases substantially during the
outburst. As an example, the change in the strength of the Balmer jump from
July 1994 to December 1995 can be seen in Fig. 4 of Viotti \etal (1996).

\subsection{Comparison with other supersoft X-ray sources}

Supersoft X-ray sources (SSS) are characterized by very soft X-ray radiation
of high luminosity. The \ros\, spectra are well described by
blackbody emission at a temperature of about kT$\approx$ 25--40 eV
and a luminosity close to the Eddington limit (Greiner \etal 1991, Heise 
\etal 1994). After the discovery of supersoft X-ray sources with \ein\,
observations, the \ros\, satellite has discovered more than a dozen new SSS.
Most of these have been observed in nearby galaxies (see Greiner (1996) for
a recent compilation).
Among the optically identified objects there are several different types 
of objects: close binaries like the prototype CAL 83 (Long \etal 1981),
novae, planetary nebulae, and symbiotic systems. 

In addition to AG Dra, two other X-ray luminous symbiotic systems with 
supersoft X-ray spectra are known (RR Tel, SMC\,3), both being symbiotic novae.
RR Tel was shown to exhibit a similar soft X-ray spectrum (Jordan \etal 1994)
and luminosity (M\"urset \& Nussbaumer 1994). RR Tel is one of the only 
seven known symbiotic nova systems. It went into outburst in 1945 with a 
brightness increase of 7$^{\rm m}$, and since then declined only slowly.
SMC\,3 (= RX J0048.4--7332) in the Small Magellanic Cloud was found to be
supersoft in X-rays by Kahabka \etal (1994), and has been observed to be
in optical outburst in 1981 (Morgan 1992). Simultaneous fitting of the \ros\, 
and UV data was possible only after inclusion of a wind mass loss from the
hot component, and gives a temperature above 260\,000 K (Jordan \etal 1996).
Symbiotic novae are generally believed to be due to a thermonuclear outburst 
after the compact object has accreted enough material from the (wind of the)
companion or an accretion disk. 

The similarity in the quiescent X-ray properties of AG Dra to those of the 
symbiotic novae RR Tel and SMC\,3 might support the speculation 
that AG Dra is a symbiotic nova in the post-outburst stage for which 
the turn-on is not documented. We have checked some early records such as 
the Bonner Durchmusterung, but always find AG Dra at the 10--11\m\, intensity 
level. Thus, if AG Dra should be a symbiotic nova, its hypothetical turn on 
would have occurred before 1855. We note, however, that there are a number
of observational differences between AG Dra and RR Tel which would be 
difficult to understand if AG Dra were a symbiotic nova.

If AG Dra is not a symbiotic nova, it would be the first wide-binary supersoft
source (as opposed to the classical close-binary supersoft sources) the 
existence of which has been predicted recently (Di\,Stefano \etal 1996). These 
systems are believed to have donor companions more massive than the accreting 
white dwarf which makes the mass transfer unstable on a thermal timescale.

\section{Summary}

The major results and conclusions of the present paper can be summarised as
follows:
\begin{itemize}
\vspace{-0.1cm}
\item The X-ray spectrum in quiescence is very soft, with a blackbody
temperature of about 14.5$\pm$1 eV. 
\item The quiescent bolometric luminosity of the hot component in the AG Dra 
binary system is (9.5$\pm$1.5)$\times$10$^{36}$ (D/2.5 kpc)$^2$ erg/s, thus
suggesting stable surface hydrogen burning in quiescence.
Adopting a distance of AG Dra of 2.5 kpc,
the X-ray luminosity suggests a low-mass white dwarf (M$<$0.6 \msun).
\item In order to sustain the high luminosity the cool companion in AG Dra
 either has to have a wind mass loss substantially larger than usual K giants
or is required to 
fill its Roche lobe, and consequently has to be brighter than a usual K giant.
The mass of the cool component should be smaller than $\approx$2 \msun\, for
the given orbital parameters and the mass ratio would be 2--5.
\item The monitoring of AG Dra at X-rays and UV wavelengths did not yield any
hints for the predicted eclipse during the times of the U light minima.
\item The recent optical outbursts of AG Dra have given us the rare opportunity
to study the rapid evolution of the X-ray and UV emission with time.
During the optical outburst in 1994 the UV continuum increased by a factor 
of 10, the UV line intensity by a factor of 2 (see Fig. \ref{iue1}), 
and the X-ray intensity dropped by at least a factor of 100 (see Fig. 
\ref{light}). There is no substantial time lag between the variations in the 
different energy bands compared to the optical variations.
\item There is no hint for an increase of the absorbing column during the one 
ROSAT PSPC X-ray observation (with spectral resolution) performed during the 
decline phase
of the 1994 optical outburst. Instead, a temperature decrease is consistent 
with the X-ray data which is also supported by the \iue\, spectral results.
\item Modelling the X-ray intensity drop by a slowly expanding white
dwarf with concordant cooling, we find that the accretion rate 
has to rise only slightly above $\dot M_{\rm RG}$. Accordingly, the white dwarf
expands to approximately its double size within the about three months
rise of the optical outburst. The cooling during this expansion is moderate: 
the temperature decreases by only about 35\%.
\item The UV continuum emission during quiescence consists of two components
which both match perfectly to the neighbouring wavelength regions (see Fig.
\ref{qnufnu}). Shortwards of $\approx$2000\AA\, it corresponds to the tail 
of the 14--15 eV hot component as derived from the X-ray observations.
However, during the optical outbursts the UV continuum shortwards of 
$\approx$2000\AA\, does not correspond to the tail of the cooling hot component 
(9--11 eV), but instead is completely dominated by another emission mechanism, 
possibly a recombination continuum in the wind zone around the hot component.
\item AG Dra could either be a symbiotic nova for which the turn on would
have occurred before 1855, or the first example of the wide-binary 
supersoft source class.

\end{itemize}

\begin{acknowledgements}
We are grateful to  J. Tr\"umper and W. Wamsteker for granting
the numerous target of opportunity observations with \ros\, and \iue.
We thank Janet A. Mattei for providing recent AAVSO observations of AG Dra,
M. Friedjung for discussions on AG Dra properties and D. Sch\"onberner and 
T. Bl\"ocker for details on the expansion/contraction rate of small core-mass 
stars.  We also thank the
referee J. Mikolajewska for a careful reading of the manuscript and helpful
comments. JG is supported by the Deutsche Agentur f\"ur
Raumfahrtangelegenheiten (DARA) GmbH under contract FKZ 50 OR 9201.
RV is partially supported by the Italian Space Agency under contract
ASI 94-RS-59 and is grateful to Dr. Willem Wamsteker for hospitality at the
IUE Observatory of ESA at Villafranca del Castillo (VILSPA).
RES thanks NASA for partial support of this research through grants NAG5-2103
(IUE) and NAG5-2094 (ROSAT) to the University of Denver. The \ros\, project is 
supported by the German Bundes\-mini\-ste\-rium f\"ur Bildung, Wissenschaft, 
Forschung und Technologie (BMBF/DARA) and the Max-Planck-Society.
\end{acknowledgements}

\end{document}